%% file: patriciateles_dpf2013.tex
\documentclass[12pt]{article}
\usepackage{graphicx}

\newcommand\pubnumber{DPF2013 - 66}
\newcommand\pubdate{\today}

\def\napoli{Centro Brasileiro de Pesquisas Físicas -- CBPF \\
Rua Dr. Xavier Sigaud 150, 22290-180, Rio de Janeiro, RJ, Brasil \\
\medskip
LHC Physics Center\\
Fermi National Accelerator Laboratory, Batavia, IL 60510 \\
\bigskip
E-mail: patricia.rebello.teles@cern.ch}
\def\support{\footnote{Work supported by Conselho Nacional de Desenvolvimento Científico e Tecnológico (CNPq) and LHC Physics Center (LPC)}}

\def\Title#1{\begin{center} {\Large #1 } \end{center}}
\def\Author#1{\begin{center}{ \sc #1} \end{center}}
\def\Address#1{\begin{center}{ \it #1} \end{center}}

\newcommand\pubblock{\rightline{\begin{tabular}{l} \pubnumber\\
         \pubdate  \end{tabular}}}
\newenvironment{Abstract}{\begin{quotation}  }{\end{quotation}}
\newenvironment{Presented}{\begin{quotation} \begin{center} 
             PRESENTED AT\end{center}\bigskip 
      \begin{center}\begin{large}}{\end{large}\end{center} \end{quotation}}

\input econfmacros.tex

\begin{document}
\begin{titlepage}
\pubblock

\vfill
\Title{Search for $WW\gamma$ and $WZ\gamma$ Production, and Anomalous Quartic Gauge Couplings in $pp$ collisions at CMS/LHC}
\vfill
\Author{ Patricia Rebello Teles\support \\
on behalf of CMS Collaboration}
\Address{\napoli}
\vfill
\begin{Abstract}

Based on a talk given at the APS DPF 2013, this proceedings deals with the study of the three gauge boson $WV\gamma$ production, containing a $W$ boson decaying leptonically, a second $V$ ($W^{\pm}$ or $Z$) boson decaying to two jets, and a photon $\gamma$, and the anomalous quartic gauge boson couplings taking into account the data collected by CMS detector at the Large Hadron Collider in 2012 at $\sqrt{s}=$8 TeV with integrated luminosity of $19.3$ fb${}^{-1}$.

\end{Abstract}
\vfill
\begin{Presented}
DPF 2013\\
The Meeting of the American Physical Society\\
Division of Particles and Fields\\
Santa Cruz, California, August 13--17, 2013\\
\end{Presented}
\vfill
\end{titlepage}
\def\thefootnote{\fnsymbol{footnote}}
\setcounter{footnote}{0}

\section{Introduction}

\hspace{0.6cm}Aiming to observe the three gauge boson production by extending the precisely measured diboson production analysis at CMS~\cite{cms,diboson} with energetic photons, in order to model the electromagnetic radiation in the $WW$ and $WZ$ pair creation processes, we notice that these channels are also suitable for the SM quartic interactions as well as sensitive to deviations of them.

Due to its non-abelian $SU(2)_{L}\otimes U(1)_{Y}$ gauge symmetry structure, from the SM naturally emerges the quartic $WWWW$, $WWZZ$, $WWZ\gamma$ and $WW\gamma\gamma$ vertices. The investigation of them is expected to play an important role at LHC energies, mainly because they have a great potencial to explore possible new physics expressed in a model independent way by high-dimensional effective operators~\cite{operators} leading to anomalous quartic gauge couplings (aQGC). Recently the anomalous $WW\gamma \gamma$ vertex was constrained through the exclusive $WW$ production~\cite{fsq}. 

This analysis~\cite{smp} is focused on $WV\gamma$ production in the semi-leptonic final state which includes $W(\to l \nu_{l}) W(\to jj)\gamma$ and  $W(\to l \nu_{l}) Z(\to jj)\gamma$ processes and was chosen due to the higher branching ratio compared to the fully leptonic mode. Since these two production reactions cannot be differentiated due to the detector di-jet mass resolution close to the mass difference between $W$ and $Z$ bosons, and the semi-leptonic final state has a common dominant background, that is $W(\to l \nu_{l})\gamma+\mbox{jets}$, then we have treated both channels as combined signal in this analysis. 

\section{Theory}

\hspace{0.6cm}Anomalous vertices may be associated either to dimension 6 or to dimension 8 effective operators, which emerge naturally from convenient gauge symmetry realizations in a model independent way~\cite{operators}.  

Concerning the anomalous quartic vertices involving photons, as $WW\gamma \gamma$ and $WWZ\gamma$, the effective operators tested in this analysis can be written as
\begin{equation}
{\cal L}_{AQGC} = \frac{a_0^W}{4 g^2} {\cal W}_{0}^{\gamma} + \frac{a_c^W}{4 g^2} {\cal W}_{c}^{\gamma} + \sum_i k_i^{W} {\cal W}_{i}^{Z} + {\cal L}_{T,0}
\label{lagrangian}
\end{equation}
where the dimension 6, $a_{0,C}^{W}$ and $\kappa_{0,C}^{W}$, parameters are associated with the $WW\gamma\gamma$ and $WWZ\gamma$ vertices respectively, while the parameter $f_{T,0}$, from $L_{T,0}$ dimension 8 operator, is associated with both. 

\begin{figure}[htb]
\centering
\includegraphics[height=2.1in]{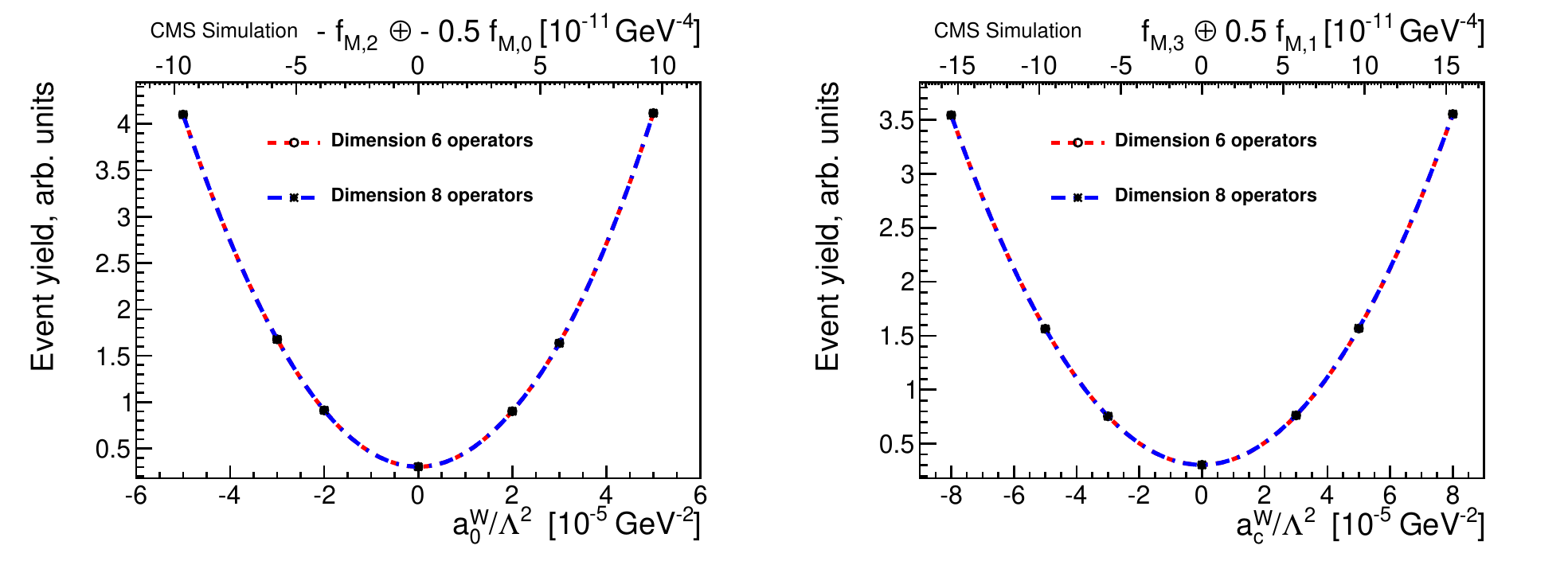}
\caption{Dimension 8 parameters $f_{M,i}$ can be associated to dimension 6 $a_{0,C}^{W}$ through simple linear transformations.}
\label{comp628}
\end{figure}

Moreover, some dimension 6 and dimension 8 operators are similar, as we can see at Fig.~\ref{comp628}, and can be related through the linear transformations
\[ \frac{f_{M0}}{\Lambda^{4}}=\frac{g^{2}}{4M_{W}^{4} s^{2}_{W} }\;\frac{k_{0}^{w}}{\Lambda^{2}};\;\frac{f_{M2}}{\Lambda^{4}} = \frac{g^{\prime 2}}{2M_{W}^{4}s^{2}_{W}}\;\frac{k_{0}^{b}}{\Lambda^{2}};\;\frac{f_{M1}}{\Lambda^{4}}=\frac{g^{2}}{2M_{W}^{4}s^{2}_{W}}\;\frac{k_{C}^{w}}{\Lambda^{2}};\;\frac{f_{M3}}{\Lambda^{4}} = \frac{g^{\prime 2}}{M_{W}^{4}s^{2}_{W}}\;\frac{k_{C}^{b}}{\Lambda^{2}}
\]
being also valid the relation 
\[a_{0,C}^{W}=4g^2(k_{0,C}^{w}+k_{0,C}^{b}+k_{0,C}^{m}).\]

\section{Selection Criteria and Simulation}

\hspace{0.6cm}For the SM $WV\gamma$ search, a cut and count approach was adopted, based on the selection criteria described here.

We have used MADGRAPH 5.1.3.22~\cite{generators}, adopting CTEQ6L1 as parton distribution function, to generate leading order (LO) events samples for signal and backgrounds. Single top samples were generated with POWHEG~\cite{generators}. A summary of the contributing processes is given in Table~\ref{XS}. 

\begin{table}[t]
\begin{center}
\begin{tabular}{|c|c|c|}
\hline
Process &  shape modeling & cross section[pb] \\ 
\hline
\hline
SM $WW\gamma$ & MC & (NLO) 0.0896 $\pm$ 0.0213 \\
SM $WZ\gamma$ & MC & (NLO) 0.0121 $\pm$ 0.0029 \\\hline
$W\gamma$+Jets & MC & (data) 10.872 $\pm$ 0.087 \\
jet $\to$ $\gamma$ & data & data \\
$Z\gamma$+Jets & MC & (LO) 0.632 $\pm$ 0.126 \\
$t\bar{t}\gamma$ & MC & (LO) 0.615 $\pm$ 0.123 \\
Single Top + $\gamma$ (inclusive) & MC & (NLO) 0.310 $\pm$ 0.011 \\ \hline
\end{tabular}
\caption{Summary of the SM processes adopted in this analysis. NLO cross section assuming $p_{T}^{\gamma}\;>$ 10 GeV and $|\eta^{\gamma}|\; <$ 2.5}
\label{XS}
\end{center}
\end{table}

All LO samples were matched to parton showers from PYTHIA 6.426~\cite{pythia}. To derive the corresponding K-factors for three boson production ($K_{WV\gamma}=2.1$) and aQGC ($K_{aQGC}=1.2$) we have used de aMC@NLO~\cite{generators}. For each aQGC parameter, several samples were generated with different values of the parameters, maintaining all other parameters equal to zero. 

The leptonic W gauge boson includes a requirement of transverse mass ($M_T^W>30$~GeV), with either one electron ($E_T^{e} >30$~GeV, $|\eta^{e}| <2.5$, excluding $1.44 < |\eta^{e}| < 1.57$) or one muon
($p_T^{\mu} >25$~GeV, $|\eta^{\mu}| < 2.1$) in the final state. Events with additional leptons are vetoed to reduce backgrounds with di- and tri-lepton final states. The neutrino induces a selection requirement of $\ensuremath{{E\!\!\!/}_{\!\mathrm{T}}} >35$~GeV.  

The two most energetic jet candidates are required to satisfy $p_T^{j} >30$~GeV and $|\eta^{j}| <2.4$. The photon candidate must satisfy $E_T^{\gamma} >30$~GeV and $|\eta^{\gamma}| <1.44$.  The data were collected with single-lepton triggers using $p_T$ thresholds of 24-30~GeV for muons and 27-32~GeV for electrons.

The azimuthal separation $\Delta\phi$ between the leading jet and the $\ensuremath{{E\!\!\!/}_{\!\mathrm{T}}}$ direction, have to be larger than 0.4 to reduce mismeasured $\ensuremath{{E\!\!\!/}_{\!\mathrm{T}}}$. To reduce W$\gamma$+jets background events, a di-jet invariant mass window of $70< m_{jj} < 100$~GeV, and a separation between the jets of $|\Delta\eta_{jj}| <1.4$, were imposed. More details, as well as a summary of all the contributing systematics uncertainties, is given at~\cite{smp}.

\section{Results}

\subsection{SM $WV\gamma$ Cross Section}

The $WW\gamma$ and $WZ\gamma$ cross section measurement in $pp$ collisions at $\sqrt{s}=$8 TeV is not accessible with the data collected in 2012 by the CMS detector due low statistics. In fact, after the cut \& count approach based on selection criteria described before we have observed 322 events observed (See Table~\ref{tab:evt}) against 341.5 $\pm$ 15.8 events predicted (See Table~\ref{XS}).

\begin{table}[htb]
\begin{center}
  \begin{tabular}{|c|c|c|}
  \hline
  Process  & muon channel & electron channel  \\
    & number of events & number of events  \\
  \hline
  \hline
  W$\gamma$+jets                   & 136.9 $\pm$ 3.5  $\pm$ 9.2  $\pm$ 0.0  & 101\
.6 $\pm$ 2.9 $\pm$ 8.0 $\pm$ 0.0  \\
  WV+jet, jet$\rightarrow \gamma$  & 33.1  $\pm$ 1.3  $\pm$ 4.6  $\pm$ 0.0  & 21.\
3  $\pm$ 1.0 $\pm$ 3.1 $\pm$ 0.0  \\
  MC $t\overline{t}\gamma$         & 12.5  $\pm$ 0.8  $\pm$ 2.9  $\pm$ 0.3  & 9.1\
   $\pm$ 0.7 $\pm$ 2.1 $\pm$ 0.2  \\
  MC single top                    & 2.8   $\pm$ 0.8  $\pm$ 0.2  $\pm$ 0.1  & 1.7\
   $\pm$ 0.6 $\pm$ 0.1 $\pm$ 0.0  \\
  MC Z$\gamma$+jets                & 1.7   $\pm$ 0.1  $\pm$ 0.1  $\pm$ 0.0  & 1.5\
   $\pm$ 0.1 $\pm$ 0.1 $\pm$ 0.0  \\
  multijets                        & $<$0.2$\pm$ 0.0  $\pm$ 0.1  $\pm$ 0.0  & 7.2\
   $\pm$ 3.6 $\pm$ 3.6 $\pm$ 0.0  \\
  \hline
  SM WW$\gamma$                    & 6.6   $\pm$ 0.1  $\pm$ 1.5  $\pm$ 0.2  & 5.0
   $\pm$ 0.1 $\pm$ 1.1 $\pm$ 0.1  \\
  SM WZ$\gamma$                    & 0.6   $\pm$ 0.0  $\pm$ 0.1  $\pm$ 0.0  & 0.5\
   $\pm$ 0.0 $\pm$ 0.1 $\pm$ 0.0  \\
  \hline
  Total predicted                  & 194.2 $\pm$ 3.9 $\pm$ 10.8  $\pm$ 0.6  & 147\
.9 $\pm$ 4.8 $\pm$ 9.6 $\pm$ 0.4  \\
  \hline
  \hline
  Data                             & 183                                    & 139\
     \\
  \hline
  \end{tabular}
  \end{center}
  \caption{Expected number of events per process, with statistical, systematic and luminosity uncertainties quoted.}
  \label{tab:evt}
\end{table}

Therefore it was only possible to set a one-sided upper limit on the cross section. For the amount of data presented here, we have set an upper limit of 0.24 pb at 95\% C.L. for $WV\gamma$ with photon $p_{T}>10$~GeV, which corresponds to 3.4 times the SM prediction.

\begin{table}[htb] 
\begin{center}
  \begin{tabular}{|c|c|}
  \hline
  Observed Limits & Expected Limits \\
  \hline
  \hline
  -21 (TeV$^{-2}$) $<$ $a_{0}^{W}/\Lambda^{2}$ $<$ 20 (TeV$^{-2}$)  & -24 (TeV$^{\
-2}$) $<$ $a_{0}^{W}/\Lambda^{2}$ $<$ 23 (TeV$^{-2}$) \\
  -34 (TeV$^{-2}$) $<$ $a_{C}^{W}/\Lambda^{2}$ $<$ 32 (TeV$^{-2}$)  & -37 (TeV$^{\
-2}$) $<$ $a_{C}^{W}/\Lambda^{2}$ $<$ 34 (TeV$^{-2}$) \\
  -25 (TeV$^{-4}$) $<$ $f_{T,0}/\Lambda^{4}$ $<$ 24 (TeV$^{-4}$)  & -27 (TeV$^{-4\
}$) $<$ $f_{T,0}/\Lambda^{4}$ $<$ 27 (TeV$^{-4}$) \\
  -12 (TeV$^{-2}$) $<$ $\kappa_{0}^{W}/\Lambda^{2}$ $<$ 10 (TeV$^{-2}$)  & -12 (T\
eV$^{-2}$) $<$ $\kappa_{0}^{W}/\Lambda^{2}$ $<$ 12 (TeV$^{-2}$) \\
  -18 (TeV$^{-2}$) $<$ $\kappa_{C}^{W}/\Lambda^{2}$ $<$ 17 (TeV$^{-2}$)  & -19 (T\
eV$^{-2}$) $<$ $\kappa_{C}^{W}/\Lambda^{2}$ $<$ 18 (TeV$^{-2}$) \\
  \hline
  \end{tabular}
  \end{center}
\caption{95\% C.L. shape-based exclusion limits listed for both the muon and el\
ectron channels of each aQGC parameter using
photon $p_{T}$.}
  \label{tab:limit_values_noMVA} \end{table}

\subsection{Exclusion Limits for the Anomalous Quartic Gauge Couplings}

\begin{figure}[htb]
\centering
\includegraphics[height=3in]{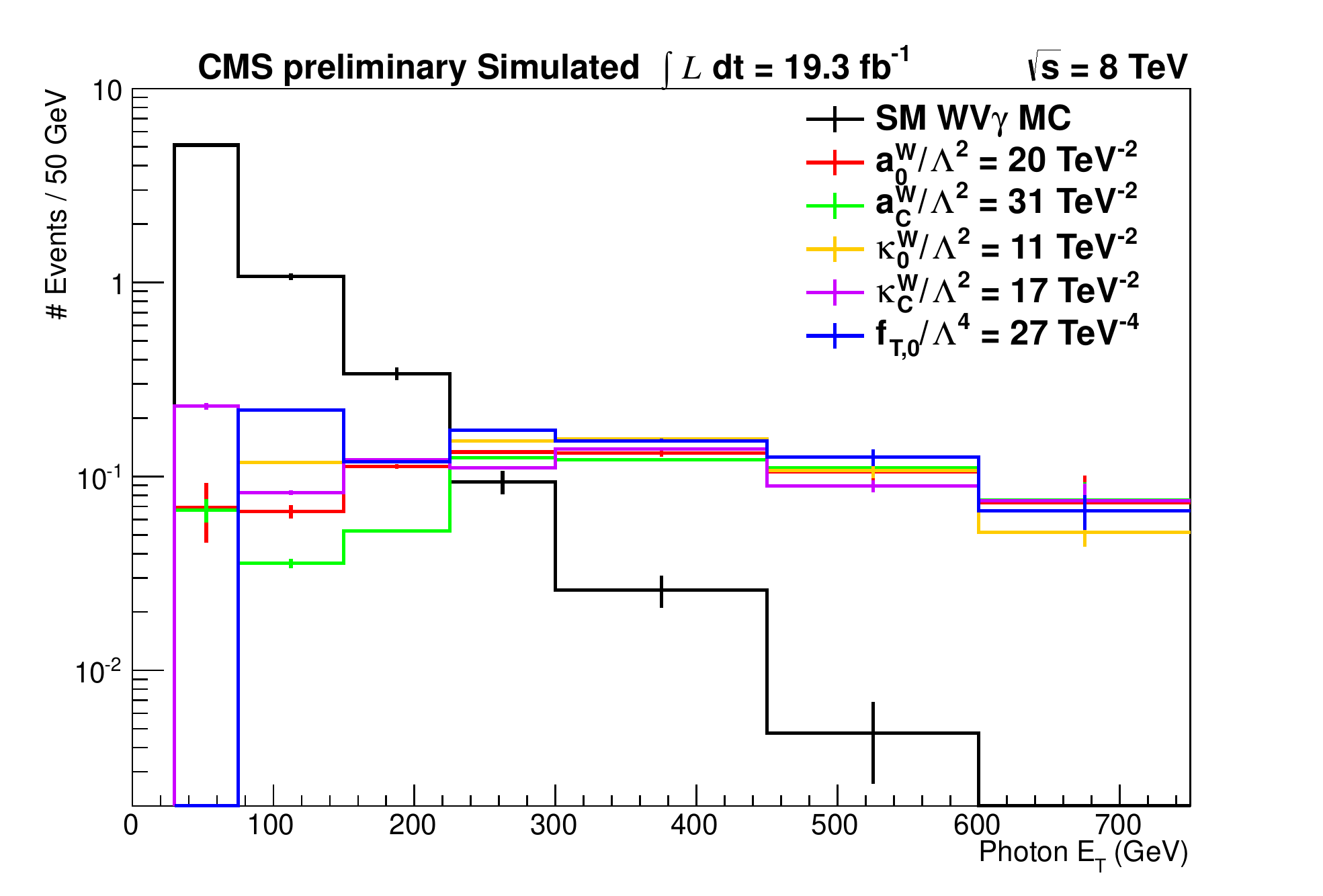}
\caption{Plot of photon $E_T$ distributions after all selections for the muon channel for the SM prediction and for the aQGC signals for the five parameters. The distributions are similar for the electron channel.}
\label{mu_limit_input}
\end{figure}

The photon $p_T$ distributions, segregated by lepton flavor, was used after all selection criteria as the observable to set limits on the aQGC parameters. See Fig.~\ref{mu_limit_input} which show the excess of events profile for the muon channel for given values of aQGC parameters. 

Finally, no evidence of anomalous $WW\gamma \gamma$ and $WWZ\gamma$ quartic gauge
couplings was found and we have computed exclusion limits (See Tables \ref{tab:limit_values_noMVA} and \ref{tab:limit_values_noMVA_dim8}) for the several aQGC parameters at the 95\% C.L. 

\begin{table}[htb]
\begin{center}
  \begin{tabular}{|c|c|}
  \hline
  Observed Limits & Expected Limits \\
  \hline
  \hline
  -77 (TeV$^{-4}$) $<$ $f_{M,0}/\Lambda^{4}$ $<$ 81 (TeV$^{-4}$)  & -89 (TeV$^{-4}$) $<$ $f_{M,0}/\Lambda^{4}$ $<$ 93 (TeV$^{-4}$) \\
  -131 (TeV$^{-4}$) $<$ $f_{M,1}/\Lambda^{4}$ $<$ 123 (TeV$^{-4}$)    & -143  (TeV$^{-4}$) $<$ $f_{M,1}/\Lambda^{4}$ $<$ 131  (TeV$^{-4}$) \\
  -39 (TeV$^{-4}$) $<$ $f_{M,2}/\Lambda^{4}$ $<$ 40 (TeV$^{-4}$)  & -44 (TeV$^{-4}$) $<$ $f_{M,2}/\Lambda^{4}$ $<$ 46 (TeV$^{-4}$) \\
  -66 (TeV$^{-4}$) $<$ $f_{M,3}/\Lambda^{4}$ $<$ 62 (TeV$^{-4}$)    & -71  (TeV$^{-4}$) $<$ $f_{M,3}/\Lambda^{4}$ $<$ 66  (TeV$^{-4}$) \\
  \hline
  \end{tabular}
  \end{center}
  \caption{95\% C.L. shape-based exclusion limits listed for both the muon and electron channels of each dimension 8 aQGC parameter, using photon PT as the observable.}
  \label{tab:limit_values_noMVA_dim8} \end{table}

A small assymmetry in the limits is expected due to the interference between the SM and aQGC processes. These are the first ever limits on dimension 8 $f_{T,0}$ and dimension 6 CP-conserving couplings $\kappa_{0,C}^{W}$.  

\begin{figure}[htb!]
\centering
\includegraphics[height=4.1in]{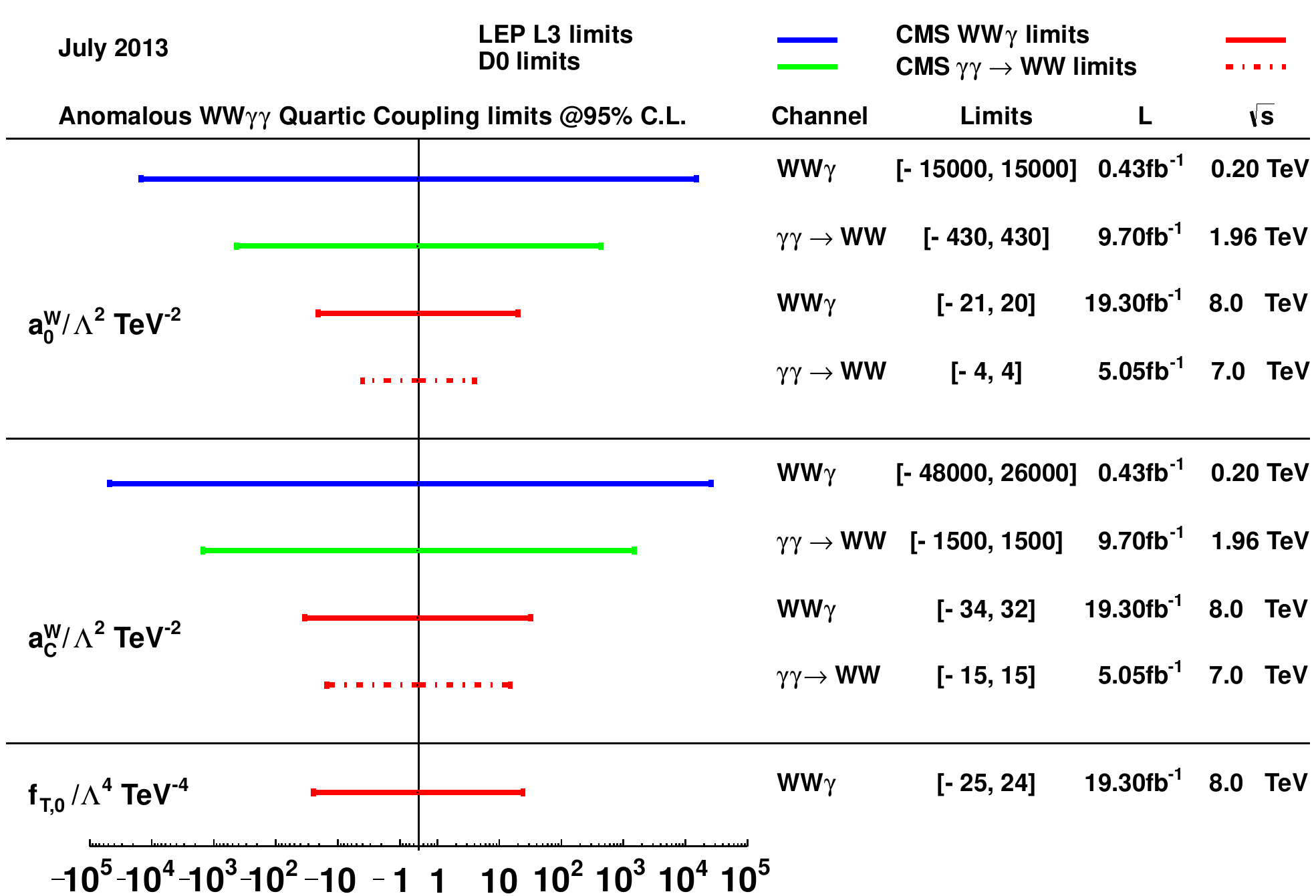}
\caption{Plot comparing the previous limits with limits from LEP, Tevatron, and other CMS measurements.}
\label{comp}
\end{figure}

A comparison of several existing limits on the $WW\gamma\gamma$ aQGC parameter is shown on Figure~\ref{comp}. The CMS limits are orders of magnitude more stringent than the best limits obtained at LEP and Tevatron.

\section*{Acknowledgments}
The author is deeply grateful to LPC Physics Center for the hospitality and support during the essential steps of this investigation.

\end{document}

%% file: econfmacros.tex
\def\beq{\begin{equation}}
\def\eeq#1{\label{#1}\end{equation}}
\def\eeqn{\end{equation}}

\def\beqa{\begin{eqnarray}}
\def\eeqa#1{\label{#1}\end{eqnarray}}
\def\eeqan{\end{eqnarray}}

\let\bar=\overbar

\def\Dslash{\not{\hbox{\kern-4pt $D$}}}
\def\dslash{\not{\hbox{\kern-2pt $\del$}}}

\def\msb{{\bar{\ssstyle M \kern -1pt S}}}

